\documentclass[
  twocolumn,
  showpacs,
  amsmath,
  amssymb,
  superscriptaddress,
  floatfix
]{revtex4-1}

\usepackage{latexsym, amsmath, amssymb, mathrsfs, graphicx}
\usepackage{color}
\usepackage{tikz}
\usepackage{booktabs}
\usepackage{array}
\usepackage{multirow}
\usepackage{hyperref}
\definecolor{linkcolour}{rgb}{0,0.2,0.6}
\hypersetup{colorlinks,breaklinks,
            linkcolor=linkcolour,citecolor=linkcolour,
            filecolor=linkcolour, urlcolor=linkcolour}

\newcommand{\ket}[1]{\left| #1 \right>}
\newcommand{\bra}[1]{\left< #1 \right|}

\newcommand{\gs}{\ket{0}_N}
\newcommand{\gsb}{{}_N\bra{0}}

\newcommand{\smfrac}[2]{\genfrac{}{}{}{1}{#1}{#2}}

\DeclareMathOperator{\Pf}{Pf}

\newcommand{\Zrec}{Z_{\mathcal{R}}(L,L')}
\newcommand{\Arec}{A_{\mathcal{R}}(\tau)}

\begin{document}

\title{Conformal boundary state for the rectangular geometry}

\author{R.~Bondesan}
\affiliation{Institute de Physique Th\'eorique, CEA Saclay, F-91191
  Gif-sur-Yvette, France}
\affiliation{LPTENS, \'Ecole Normale Sup\'erieure, 24 rue Lhomond,
  75231 Paris, France}
\affiliation{Institut Henri Poincar\'e, 11 rue Pierre et Marie Curie,
  75231 Paris, France}

\author{J.~Dubail}
\affiliation{Department of Physics, Yale University, P.O. Box 208120,
  New Haven, CT 06520-8120, USA}

\author{J.~L.~Jacobsen}
\affiliation{LPTENS, \'Ecole Normale Sup\'erieure, 24 rue Lhomond,
  75231 Paris, France}
\affiliation{Universit\'e Pierre et Marie Curie, 4 place Jussieu,
  75252 Paris, France}
\affiliation{Institut Henri Poincar\'e, 11 rue Pierre et Marie Curie,
  75231 Paris, France}

\author{H.~Saleur}
\affiliation{Institute de Physique Th\'eorique, CEA Saclay, F-91191 Gif-sur-Yvette, France}
\affiliation{Physics Department, USC, Los Angeles, CA 90089-0484, USA}
\affiliation{Institut Henri Poincar\'e, 11 rue Pierre et Marie Curie,
  75231 Paris, France}

 \begin{abstract}
   \noindent 
   We discuss conformal field theories (CFTs)
  in rectangular geometries, and develop a formalism that involves
 a conformal boundary state for the 
 $1+1$d open system. 
   We focus on the case of homogeneous boundary conditions (no insertion of a boundary condition changing operator),
   for which we derive an explicit expression of the associated 
   boundary state, valid
   for any arbitrary CFT.
   We check the validity of our solution, comparing it with
   known results for partition functions,
   numerical simulations of lattice discretizations, and coherent state
   expressions for free theories.
 \end{abstract}


\maketitle

\section{Introduction}
\label{sec:intro}

Boundary conformal field theory (BCFT)  is a subject whose importance has grown over the years, both on the formal and on the applied side. It does for instance play a fundamental role in the axiomatization of conformal field theory (CFT) \cite{Zuber}, 
in our growing understanding of logarithmic conformal field theory \cite{Runkel,Read2007}, or in the relationship between conformal field theory and the Schramm Loewner Evolution formalism \cite{BauerBernard}.  It is also central to our understanding of the Kondo effect \cite{Affleck}, 
of the physics of quantum impurities or the Fermi edge singularity \cite{AffleckLudwig}, and, more recently, of local and global quenches in one dimensional quantum systems 
\cite{CardyCalabrese, DubailStephan}.

Two geometries are most naturally used in BCFT. One of them---more natural from the point of view of  Euclidian field theory or statistical mechanics---is simply the upper half plane, with the theory defined \emph{e.~g.~}only for $\hbox{Im }z\geq 0$, together with some boundary conditions at $\hbox{Im }z=0$. The other involves a cylinder (sometimes considered instead as an annulus), and typically describes the physics of an open $1D$ quantum system at finite temperature. 

Of course, there are variants of these geometries. For instance, the physics of the cylinder can be described with an (imaginary) time evolution along the axis. In this case, the $1D$ quantum system is closed, and the presence of the boundary is encoded in a boundary state. Such states have given rise to many developments, in the context of efforts to classify all conformal boundary conditions in particular. 

We shall be concerned in this paper mostly with systems on a rectangle, which thus have trivial topology, sharp corners, and, in general, four different boundary conditions on their four edges. While in principle this situation can be tackled via conformal mappings of the half plane (see below), few of its features have actually been studied in detail. The situation is, on the other hand, clearly interesting in its own right. The rectangle geometry is, for instance, the natural one to consider in the case of quenches for $1D$ quantum systems with open boundaries. In the $2D$ point of view, it is the simplest geometry to study transport properties of network models for Anderson localization. More fundamentally, the rectangle provides a natural way to study and interpret the  conformal blocks of four point functions by inserting four different fields at the four corners. This is particularly useful for instance to connect geometrical correlators in the Self Avoiding Walk
 (SAW) problem to CFT, in particular in the logarithmic (indecomposable) case. 

We will focus on these and other geometrical features in subsequent and rather technical work to appear soon. The present paper is devoted to the exploration of the simplest---and yet very rich---aspects of this problem related with the boundary state for a theory defined on a segment. 

\section{Gluing condition}
\label{sec:gluing}

For a CFT in the complex plane $\mathbb{C}$, one can implement
a boundary with the requirement that there is no energy
flow across it. In other words, one of the components of
the stress-tensor vanishes along
the boundary: $T_{\parallel \perp}=0$. In holomorphic/anti-holomorphic
components this gives the constraint $T(z) = \bar{T}(\bar{z})$ on the boundary.
In radial quantization, a circular boundary at radius $|z|=1$ can be encoded in the form of
a boundary state $\left| B^p \right>$. Here the subscript $p$ stands for
periodic since the boundary is defined on a circle. The
constraint is
\begin{equation}
  \label{eq:gluing_cyl}
  \left(L_n - \bar{L}_{-n}\right) |B^{p}\rangle = 0 \, ,
\end{equation}
which is usually referred to as the conformal invariance of the boundary condition,
or {\it gluing condition}, as it glues the modes of the chiral part of the CFT
with the anti-chiral ones. A basis of solutions of the linear system of equations (\ref{eq:gluing_cyl}) is given by the   so-called Ishibashi states, which are particular combinations of left-right symmetric Virasoro descendants 
of primary fields. Of course, to obtain the allowed boundary states  $|B^{p}\rangle$ themselves, more conditions have to be implemented  
(see for example \cite{Zuber,Affleck} for reviews).

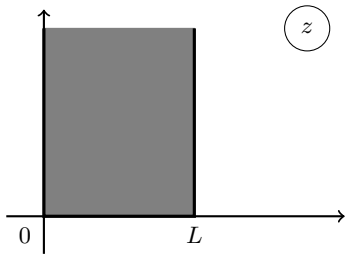
\begin{figure}[hpt]
\begin{center}
\begin{tikzpicture}[scale=1]
  \filldraw[gray] (0,0) rectangle (2,2.5);  
  \draw [->,thick] (-0.5,0) -- (4,0) ;
  \draw [->,thick] (0,-0.5) -- (0,2.75) ;
  \draw [very thick] (2,2.5) -- (2,0) -- (0,0) -- (0,2.5) ;
  \node at (-0.25,-0.25) {$0$};
  \node at (2,-0.25) {$L$};
  \draw (3.5,2.5) circle (0.3);
  \node at (3.5,2.5) {$z$};
\end{tikzpicture}
\end{center}
\caption{The semi-rectangular geometry defining the boundary state.
}
\label{fig:bstate}
\end{figure}

In this paper, we shall extend the formalism of boundary {\it states} 
to the case of an open system. Namely,
we are now considering the CFT on a strip of width $L$, and we want
to find a state in the theory that encodes the  boundary perpendicular
to the direction of the strip (see Fig.~\ref{fig:bstate}). In the
$1+1$ Hamiltonian description of the CFT, the
boundary state now lives in the Hilbert space of the $1$d theory
defined on a segment instead of a circle.  To proceed, we choose coordinates $z=x+iy$ in
the plane and consider the semi-rectangular region $0\le x\le L, y\ge
0$ of Fig.~\ref{fig:bstate}, and derive the gluing condition for the
stress-energy tensor in this geometry (for analogous discussions see
\cite{Imamura2006,Imamura2008,Yin2007}).
Conformal invariance of the boundary implies then $T_{xy}=0$ on every side.
This is realized at the boundaries $z=iy$ and $z=L+iy$, $y>0$, by
setting
\begin{equation}
  \label{eq:TTbar}
  \bar{T}(\bar{z}) = T(2L-\bar{z})\, ,
\end{equation}
and imposing periodic boundary conditions in the $x$ direction of
period $2L$: $T(z+2L)=T(z)$ (same for $\bar{T}$).

Care is needed to treat properly the effect of corners.
The stress tensor $T$ has a singularity as its argument approaches the
corners. To see this we start more generally by the singularity in the
upper half plane when an operator of weight $h$ is inserted at the
origin. In this case the most singular term is
\begin{equation}
  \label{eq:sing_T_UHP}
  T(w)\approx \frac{h}{w^2}\, .
\end{equation}
We then fold the upper half plane by the mapping
$z=w^{1/2}$ to have a corner in $z=0$. After using the transformation
law of the stress tensor $T(z) = T(w) ({\rm d}w/{\rm d}z)^2+(c/12)
\{w;z\}$, where $\{w;z\}$ is the Schwarzian derivative
\begin{equation}
  \label{eq:schwarz_der}
  \{w;z\}\equiv \frac{{\rm d}^3w/{\rm d}z^3}{{\rm d}w/{\rm d}z}   
  -\frac{3}{2} \left(\frac{{\rm d}^2w/{\rm d}z^2}
    {{\rm d}w/{\rm d}z} \right)^2\, ,
\end{equation}
we get
\begin{equation}
  \label{eq:sing_T_corner}
  T(z)\approx \left(4h-\frac{c}{8}\right) \frac{1}{z^2}\, .
\end{equation}
Also if $h=0$, at a corner there is an anomaly, which reflects itself
in a non-trivial scaling dependence of physical quantities \cite{Cardy1988,Vernier2011}.

As a consequence the condition on $T$ defining the boundary state
$|B^o \rangle$ at $y=0$ is
\begin{equation}
  \label{eq:}
  \left(T(x)-T(-x)+4\pi i\left(\tilde{h}_l\delta'(x) + \tilde{h}_r\delta'(x-L)
    \right) \right)
  |B^o\rangle = 0 \, ,
\end{equation}
where we have used
\begin{equation}
  \label{eq:deltap}
  \frac{1}{(x-i\epsilon)^2}-\frac{1}{(x+i\epsilon)^2}
  \approx
  \frac{4i\epsilon x}{(x^2+\epsilon^2)^2}\to -2\pi i\delta'(x)\, ,
\end{equation}
and defined the ``effective conformal weight'' at the corners
\begin{equation}
  \label{eq:htilde}
  \tilde{h}:=2 h-\frac{c}{16} \, ,
\end{equation}
$h_{l}$ ($h_r$) being the weight of the operator inserted at the left
(right) corner.  Since the system is periodic with period $2L$, we can
go to a mode expansion
\begin{equation}
  \label{eq:T_modes_circle}
  T(x)=-\frac{\pi^2}{L^2}\left(\sum_n L_n e^{-i\pi n x/L}-\frac{c}{24}\right)\, ,
\end{equation}
to get the gluing condition (for $n\in \mathbb{Z}^{>0}$) 
\begin{equation}
  \label{eq:gluing_rect}
  \left(L_n - L_{-n} -2n\left(\tilde{h}_l + (-)^n \tilde{h}_r\right) \right)
  |B^o\rangle = 0 \, .
\end{equation}
Note in particular that in this case one identifies
not only $L_n$ with $\bar{L}_{-n}$ but also $L_n$ with $L_{-n}$, since
the semi-rectangular geometry is obtained by two consecutive folding
of the plane.

Calling the boundary conditions on the sides of the semi-rectangle
$a,b,c$, and the boundary condition changing operators sitting at the
corners $\phi_i^{a|b}(0)$ and $\phi_j^{b|c}(L)$ (with
opportune labels $i,j$), the boundary state $|B^o\rangle$ will
generically live in a direct sum of vector spaces determined by the
fusion $\phi_i^{a|b}\otimes \phi_j^{b|c}$. In this paper we 
consider the case of homogeneous boundary conditions $a=b=c$,
where the identity operator sits at the corners ($h_l=h_r=0$),
so that equation \eqref{eq:gluing_rect} becomes
\begin{equation}
  \label{eq:gluing_rect0}
  \left(L_n - L_{-n} + \frac{n c}{8} \left[1 + (-1)^n \right] \right)
  |B^o\rangle = 0 \, ,
\end{equation}
to be solved within the Verma module of the vacuum $\left|0 \right>$
of the CFT.  The solution of the homogeneous case will then be of the
form $G\left|0\right>$, with $G$ a certain expression in terms of the  Virasoro generators $L_{-n}$'s.
When boundary condition changing operators are present, 
the boundary state will simply be \,
$G\;\phi_i^{a|b}(w_1)\phi_j^{b|c}(w_2) \left|0 \right>$, where the two points $w_1$ and $w_2$ lie on the boundary and correspond to the images of the two corners at $z=0$ and $z=L$ under a conformal mapping to the upper half plane.
Further details about this 
general situation will be discussed in a sequel \cite{Bondesan2011}.

\section{Boundary state in boundary CFT, and the rectangular bottom}
\label{sec:BstateSLE}

In this section we give the explicit form of the boundary state $\left|B^o \right>$
which solves the constraint (\ref{eq:gluing_rect0}), following the discussion in
\cite{jthesis}. Similar tricks have been used, for instance, in \cite{BauerBernard,Dubail2010}.

\subsection{CFT in the half-plane, deformations of the boundary around
  the origin}
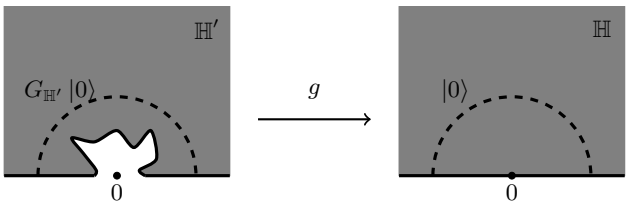
\begin{figure}
\begin{center}
\begin{tikzpicture}[scale=0.75]
	\begin{scope}
	\filldraw[gray] (-2,0) rectangle (2,3);
	\draw[very thick, white] (-0.4,0) -- (0.5,0);
	\filldraw[white,smooth] plot coordinates{(-0.4,0) (-0.37,0.1) (-0.6,0.4) (-0.8,0.67) (-0.4,0.6) (0,0.8) (0.3,0.5) (0.6,0.8) (0.7,0.3) (0.4,0.1) (0.5,0)};
	\draw[very thick,smooth] plot coordinates{(-0.4,0) (-0.37,0.1) (-0.6,0.4) (-0.8,0.67) (-0.4,0.6) (0,0.8) (0.3,0.5) (0.6,0.8) (0.7,0.3) (0.4,0.1) (0.5,0)};
	\draw[very thick] (-2,0) -- (-0.4,0) (0.5,0) -- (2,0);
	\draw (1.6,2.6) node{$\mathbb{H}'$};
	\draw[dashed, very thick] (-1.4,0) arc (180:0:1.4); 
	\filldraw (0,0) circle (1.7pt) node[below]{$0$};
	\draw (-1,1.5) node{$G_{\mathbb{H}'}\left| 0 \right>$};
	\end{scope}
	\draw[->,thick] (2.5,1) -- (4.5,1);
	\draw (3.5,1.5) node{$g$};
	\begin{scope}[xshift=7cm]
	\filldraw[gray] (-2,0) rectangle (2,3);
	\draw[very thick] (-2,0) -- (2,0);
	\draw (1.6,2.6) node{$\mathbb{H}$};
	\draw[dashed, very thick] (-1.4,0) arc (180:0:1.4); 
	\filldraw (0,0) circle (1.7pt) node[below]{$0$};
	\draw (-1,1.5) node{$\left| 0 \right>$};
	\end{scope}
\end{tikzpicture}
\end{center}
\caption{A deformation of the boundary around the origin can be
  encoded in a conformal mapping $g: \mathbb{H}' \rightarrow
  \mathbb{H}$.}
\label{fig:choux-fleur}
\end{figure}

Let us consider some boundary CFT defined in the half-plane
$\mathbb{H}$. We make use of radial quantization in $\mathbb{H}$.
The vacuum of the theory can be written formally as a path
integral
\begin{equation}
	\left| 0 \right> = \int \left[ d\phi (|z|<1, z \in \mathbb{H}) \right] \, e^{-S[\phi]} \, \left| \phi(|z|=1) \right>\, ,
\end{equation}
where $\phi(|z|=1)$ stands for the configurations of the fields of the
theory on a semi-circle centered at the origin $0$ and of radius
$1$. The weights of the different configurations is given by a Gibbs
distribution $e^{-S[\phi]}$.  In this section, we consider
deformations of the half-plane around the origin, such as the one
shown in Fig.~\ref{fig:choux-fleur}. More precisely, we take some
domain $\mathbb{H}' = \mathbb{H}\setminus K$ with $K$ a domain included in
the semi-disc of radius $1$ centered on the origin, such that
$\mathbb{H}'$ is simply connected. Our goal is to find an expression
for the new state
\begin{equation}
  \left | \mathbb{H}' \right> = \int \left[ d\phi (|z|<1, z \in \mathbb{H}') \right] \, e^{-S[\phi]} \, \left| \phi(|z|=1) \right>\, .
\end{equation}
In this paper we do not consider insertions of boundary condition
changing operators or other operators in the bulk. Therefore, we
expect $\left| \mathbb{H}' \right>$ to be given by some linear
combination of the vacuum $\left|0 \right>$ and its descendants
only. In other words, there must exist some operator $G_{\mathbb{H}'}$
built out of the Virasoro generators such that
\begin{equation}
	\left| \mathbb{H}'\right> = G_{\mathbb{H}'} \left| 0 \right>\, .
\end{equation}
Our goal is to find how to construct $G_{\mathbb{H}'}$.

By Riemann's mapping theorem, there is a conformal mapping from
$\mathbb{H}'$ onto $\mathbb{H}$. If one requires that the behavior of
$g(z)$ as $z \rightarrow \infty$ be
\begin{equation}
	g(z) = z + \frac{a_1}{z} +\frac{a_2}{z^2} + \frac{a_3}{z^3} + \dots
\end{equation} 
then the mapping $g$ is unique.

Now, let us assume that we have a continuous family of deformations of
the half-plane $\{ \mathbb{H}_t \}_{0 \leq t \leq 1}$, such that
$\mathbb{H}_0 = \mathbb{H}$ and $\mathbb{H}_1 = \mathbb{H}'$. For each
$t$ there is a mapping $g_t$ from $\mathbb{H}_t$ onto $\mathbb{H}$
with the above asymptotic behavior. The composition $g_t^{-1} \circ
g_{t+dt}$ gives us an infinitesimal mapping from $\mathbb{H}_{t+dt}$ onto
$\mathbb{H}_t$, which can be expanded as
\begin{equation}
	\label{eq:diff1}
	g_t^{-1} \circ g_{t+dt} (z) \; =\;  z \, +\, dt \left( \frac{b_1}{z} + \frac{b_2}{z^2} + \frac{b_3}{z^3} + \dots \right)
\end{equation}
This allows us to relate $\left|\mathbb{H}_{t}\right>$ and
$\left| \mathbb{H}_{t+dt} \right>$, using the definition of the
stress-tensor in boundary CFT, namely the variation of the action $S[\phi]$ under a small
transformation $z \mapsto z + \alpha(z)$ (see {\it e.g.} \cite{CardyBCFT84})
\begin{equation}
\delta S = \frac{1}{2 \pi i} \oint \alpha(z) T(z) dz + {\rm c.c.}
\end{equation}
which gives for $\alpha(z) = dt ~ (b_1/z +b_2/z^2 \dots)$
\begin{eqnarray}
  \nonumber	\left| \mathbb{H}_{t+dt} \right> &=& \int [d\phi(|z|<1, z \in \mathbb{H}_t)] (1-\delta S) e^{-S}  \left| \phi(|z|=1) \right>  \\
\nonumber  &=& \left| \mathbb{H}_{t} \right> \\
\nonumber &&-\frac{1}{2\pi i}\left( \oint \alpha(z) T(z) dz -  \oint \bar{\alpha}(\bar{z}) \bar{T}(\bar{z}) d \bar{z} \right)  \left| \mathbb{H}_{t} \right> \\
\nonumber  &=& \left| \mathbb{H}_{t} \right> - dt \left( b_1 L_{-2} + b_2 L_{-3} + b_3 L_{-4} + \dots \right) \left| \mathbb{H}_t\right>\, .
\end{eqnarray}
We arrive at the differential equation
\begin{equation}
	\frac{d }{d t} \left| \mathbb{H}_t \right> = -\left( b_1 L_{-2} + b_2 L_{-3} +b_3 L_{-4} + \dots \right) \left| \mathbb{H}_t \right>
\end{equation}
or, in terms of the operator $G_t \equiv G_{\mathbb{H}_t}$
\begin{equation}
	\label{eq:diff2}
	\frac{d}{dt} G_t = - \left(  b_1 L_{-2} + b_2 L_{-3} +b_3 L_{-4} + \dots \right)  G_t \, .
\end{equation}
Note that (\ref{eq:diff1}) and (\ref{eq:diff2}) provide us with a
differential equation which, at least formally, allows us to find the
operator $G_{\mathbb{H}'} = G_{t=1}$. This differential equation will
play an important role in the next sections.

\subsection{The boundary state}
\label{sec:the_bstate}
In general, it is not possible to find an explicit solution to the
above differential equation given some geometry $\mathbb{H}'$. For
certain cases, however, solving the equation is easy.

\begin{figure}
\begin{center}
\begin{tikzpicture}[scale=1.3]
	\filldraw[gray] (-2,0) rectangle (2,3);
	\draw[very thick] (-2,0) -- (2,0);
	\draw[very thick] (0,0) -- (0,1) node[right]{$\sqrt{2 t}$};
	\draw[very thick, dashed] (-1.4,0) arc (180:0:1.4);
\end{tikzpicture}
\end{center}
\caption{The half-plane minus a slit of height $\sqrt{2t}$, namely
  $\mathbb{H}\setminus\left[ 0, i \sqrt{2 t} \right]$, is the simplest
  example of a domain which can be encoded in a boundary state: it
  corresponds to $e^{-t L_{-2}} \left| 0 \right>$.}
\label{fig:1slit}
\end{figure}
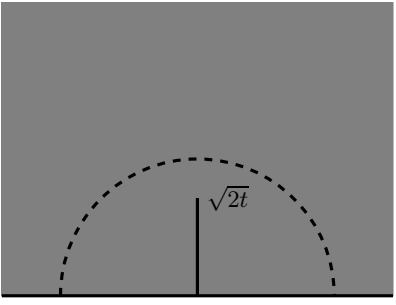

Let us start with the case of a vertical slit of height $\sqrt{2t}$ in
the half-plane, namely the domain $\mathbb{H}\setminus \left[0,i
  \sqrt{2 t}\right]$, see Fig.~\ref{fig:1slit}. It can be mapped
onto the half-plane by $g_t: z \mapsto \sqrt{z^2+2t}$. Note that
$g_t(z) = z + \frac{t}{z} +\dots$ as $z\rightarrow \infty$, so it has
the required asymptotic behavior. Differentiating this function with
respect to $t$, we get
\begin{equation}
	g_t^{-1} \circ g_{t+dt} (z) = z + \frac{dt}{z}
\end{equation}
According to the previous section, this corresponds to an operator $G_{t}$
which is given by
\begin{equation}
	\frac{d}{dt} G_t = - L_{-2} G_t
\end{equation}
with the initial condition $G_{t=0} = 1$. We thus find $G_t = e^{-t
  L_{-2}}$.  In particular, the state corresponding to a slit of
height $1$ is $e^{- \frac{1}{2} L_{-2}}$.

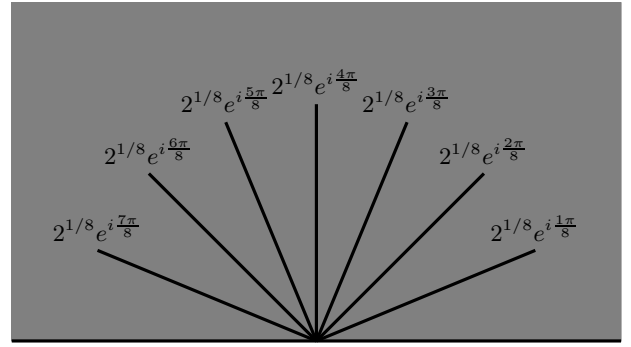
\begin{figure}
\begin{center}
\begin{tikzpicture}[scale=0.9]
	\filldraw[gray] (-4.5,0) rectangle (4.5,5);
	\draw[very thick] (-4.5,0) -- (4.5,0);
	\foreach \a in {1,2,...,7}
		\draw[very thick, rotate=22.5*\a] (0,0) -- (3.5,0) node[above]{$2^{1/8} e^{i \frac{\a \pi}{8}}$};
\end{tikzpicture}
\end{center}
\caption{The domain $\mathbb{H}_8$ is the half-plane minus $7=8-1$ slits of size $2^{1/8}$.}
\label{fig:H8}
\end{figure}

It turns out that one can easily extend this trick to a larger set of
domains with slits. Let us define the half-plane minus $k-1$ slits of
size $2^{1/k}$
\begin{equation}
	\mathbb{H}_k = 
        \mathbb{H} \setminus \left\{ z | z^k  \in \left[ -2,2\right]\right\}\, .
\end{equation}
See Fig.~\ref{fig:H8} for an illustration of the multi-slit geometry,
and note that $\mathbb{H}_1 = \mathbb{H}$. The point of this
definition is the following. We define the functions
\begin{equation}
	g_k (z) = \left( z^k+2 \right)^{1/k}
\end{equation}
and we make the observation that $g_{2^N}$ is a conformal mapping from
$\mathbb{H}_{2^N}$ onto $\mathbb{H}_{2^{N-1}}$. Moreover, these
mappings have the correct asymptotic behavior at infinity, and one
can build a family of mappings $g_{k,t} = \left(z^k +2 t\right)^{1/k}$
which gives rise to
\begin{equation}
	g_{k,t}^{-1} \circ g_{k,t+dt} (z) = z + \frac{2}{k}\frac{dt}{z^{k-1}}
\end{equation}
and therefore to a differential equation for the operator $G_{k,t}
\equiv G_{\mathbb{H}_{k,t}}$
\begin{equation}
	\frac{d}{dt} G_{k,t} = -\frac{2}{k} L_{-k} G_{k,t}
\end{equation}
which gives in the end $e^{-\frac{2 t}{k}L_{-k}}$.

As already mentioned, $g_{2^N}$ is a mapping from
$\mathbb{H}_{2^N}$ onto $\mathbb{H}_{2^{N-1}}$, so by composing
several of those, we see that $g_2 \circ g_4 \circ \dots \circ
g_{2^N}$ is a mapping from $\mathbb{H}_{2^N}$ onto the half-plane
$\mathbb{H}$. Thus, it is now straightforward to write down the
boundary state $\left|\mathbb{H}_{2^N} \right>$ with $2^N-1$ slits of
size $2^{1/2^N}$
\begin{equation}
  \label{eq:bstate_Hk}
	\left| \mathbb{H}_{2^N}\right> = e^{-\frac{1}{2^{N-1}} L_{-2^N}} \dots e^{-\frac{1}{2} L_{-4}} e^{-L_{-2}} \left| 0\right>\, .
\end{equation}

\begin{figure}
\begin{center}
\begin{tikzpicture}[scale=0.75]
	\begin{scope}
	\filldraw[gray] (-2,0) rectangle (2,3);
	\draw[very thick, white] (-1.4,0) -- (1.4,0);
	\draw[very thick,red] (-2,0) -- (-1.4,0); 
        \draw[very thick,blue] (1.4,0) -- (2,0);
	\draw (1.6,2.6) node{$\mathcal{D}$};
        \filldraw[white,smooth] (-1.4,0) arc (180:0:1.4);
	\draw[very thick] (-1.4,0) arc (180:0:1.4); 
        \node at (-1.4,-0.5) {$-1$};
        \node at (1.4,-0.5) {$1$};
	\end{scope}
        \node at (3.5,1.5) {$w=f(z)$};
	\draw[->,thick] (2.5,1) -- (4.5,1);
	\begin{scope}[xshift=7cm]
	\filldraw[gray] (-2,0) rectangle (2,3);
        \draw[very thick,red] (-2,0) -- (-1.4,0); 
        \draw[very thick,blue] (1.4,0) -- (2,0);
	\draw[very thick] (-1.4,0) -- (1.4,0);
	\draw (1.6,2.6) node{$\mathbb{H}$};
        \node at (-1.4,-0.5) {$-2$};
        \node at (1.4,-0.5) {$2$};
	\end{scope}
\end{tikzpicture}
\end{center}
\caption{Mapping of semicircular region $\cal{D}$ to the upper half plane
$\mathbb{H}$ $f(z)=z+z^{-1}$}
\label{fig:map_D_UHP}
\end{figure}
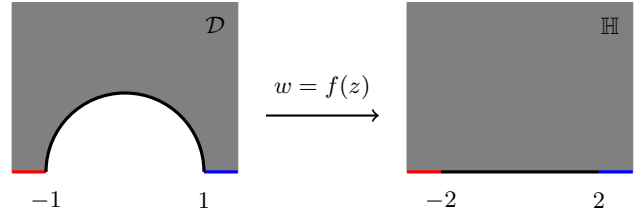

\begin{figure}
\begin{center}
\begin{tikzpicture}
	\begin{scope}[scale=0.4]
	\filldraw[gray] (-4.5,0) rectangle (4.5,5);
	\draw[very thick] (-4.5,0) -- (4.5,0);
	\foreach \a in {1,2,...,31}
		\draw[very thick, rotate=5.625*\a] (0,0) -- (3,0);
	\end{scope}
	\draw (2.4,1) node{$\underset{N\rightarrow \infty}{\longrightarrow}$};
	\begin{scope}[scale=0.4,xshift=12cm]
	\filldraw[gray] (-4.5,0) rectangle (4.5,5);
	\filldraw[white] (-3,0) arc (180:0:3);
	\draw[very thick] (-4.5,0) -- (-3,0) arc (180:0:3) -- (4.5,0);
	\end{scope}
\end{tikzpicture}
\end{center}
\caption{The boundary state $\left|B^o\right>$ is obtained in the
  $N\rightarrow \infty$ limit.}
\label{fig:limit}
\end{figure}
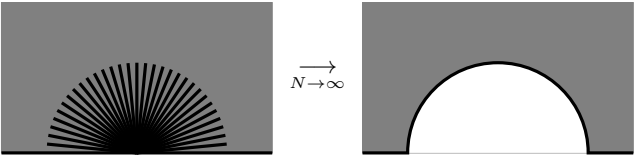

The connection between these states and the main topic of this paper
appears when one considers the limit $N \rightarrow \infty$. Indeed,
one can check that
\begin{align}
\label{eq:asymptotic}
 \nonumber 	& g_2 \circ g_4 \circ \dots \circ g_{2^N} (z)  =\\\nonumber
&\qquad = \sqrt{\sqrt{\dots \sqrt{\sqrt{z^{2^N}+2}+2}\dots+2}+2}  \\
   &\qquad =  z + \frac{1}{z} + O(z^{1-2^{N+1}})
\end{align}
when $|z| \rightarrow \infty$. We therefore have 
$\underset{N \rightarrow \infty}{\lim} g_2 \circ g_4 \circ \dots \circ g_{2^N} (z) =z +z^{-1}$ when $|z|>1$. 
The function $z \mapsto z +z^{-1}$ is a conformal mapping from
$\mathcal{D}:=\mathbb{H} \setminus \left\{z, |z| \leq 1 \right\}$
to $\mathbb{H}$, see Fig.~\ref{fig:map_D_UHP}. 
It is thus tempting to 
conclude that the boundary state $\left|B^o \right>$ which corresponds
to this domain (Fig.~\ref{fig:limit}) is \cite{jthesis}
\begin{equation}
  \label{eq:Bstate}
	\left| B^o\right> = \underset{N \rightarrow \infty}{{\rm lim}} e^{- \frac{1}{2^{N-1}} L_{-2^N}} \dots e^{-\frac{1}{2} L_{-4}} e^{-L_{-2}} \left| 0\right> \, .
\end{equation}
At first sight, this expression might look ill-defined, however it is not.
Indeed, it is obvious that, expanding the different exponentials
in powers of the Virasoro modes, one gets a finite sum at each level. For instance,
our boundary state is, up to level $4$
\begin{equation}
  \label{eq:Bstate_first_lev}
  \left| B^o\right> 
  =
  | 0\rangle
  -
  L_{-2} | 0\rangle
  -
  \frac{1}{2} L_{-4} | 0\rangle 
  +
  \frac{1}{2} L_{-2}^2 | 0\rangle + \dots \, .
\end{equation}

The foregoing argument leading to eq.~(\ref{eq:Bstate}) is of course not entirely satisfactory, since it relies on an a priori uncomfortable identification of the half disk with an infinity of slits, supported by the asymptotic behavior (\ref{eq:asymptotic}). This can hold only
 in ``some sense''---we believe, in the sense that all correlation functions in the CFT for the geometry of interest can be obtained using eq.~(\ref{eq:Bstate}). We explore this further in what follows.

\subsection{Deformed gluing conditions}
\label{sec:def_gluing}

In this section we will derive the gluing conditions satisfied by the
stress tensor in the multi-slit geometries $\mathbb{H}_{2^N}$ introduced
in section \ref{sec:the_bstate}. This will lead
to a proof of the fact that the boundary state \eqref{eq:Bstate}
satisfies the gluing condition for the semi-rectangular geometry,
eq.~\eqref{eq:gluing_rect0}.

The region $\mathbb{H}_{2^N}$ can be mapped to the upper half plane
by the conformal mapping \eqref{eq:asymptotic}, whose expansion at $z\to\infty$
is
\begin{equation}
  \label{eq:cmap_Hk_uph}
  w = f_N(z) := z + \frac{1}{z} + \sum_{p=K}^\infty \frac{\alpha_k}{z^k}\, ,
\end{equation}
where we have introduced $K:=2^{N+1}-1$. Its inverse has the following expansion
at $w\to\infty$:
\begin{equation}
  \label{eq:cmap_uph_Hk}
  z = h_N(w) := \frac{w+\sqrt{w^2-4}}{2} + 
  \sum_{p=K}^\infty \frac{\beta_k}{w^k}\, .
\end{equation}
For the general argument given below we will not need the
explicit expression of $\alpha_k$'s and $\beta_k$'s so we leave
them unspecified. We note however that closed expressions
of the above mappings can be found:
\begin{align}
   f_N(z)&=2 
   \cos\left(
     2^{-N} \arccos\left(
       \smfrac{z^{2^N}}{2}
     \right) 
   \right) \, , \\
  h_N(w)&=
  \left( 2 \cos\left(
      2^{N} \arccos\left(\smfrac{w}{2}\right)
    \right)
  \right)^{2^{-N}}\, ,
\end{align}
which eventually allow to write explicitly the coefficients appearing
above and in the next computations. Let us now derive the gluing
condition for the multi-slit geometry by using a slightly different
point of view from the one of section \ref{sec:gluing}.

If we call $X=\cdots \phi(z_1,\bar{z}_1)\cdots \psi(x)\cdots$ 
a chain of arbitrary bulk and boundary operators, by 
definition our boundary state \eqref{eq:bstate_Hk} is such that 
\begin{equation}
  \label{eq:B_vac_map}
  \left< 0\right|X  \left| \mathbb{H}_{2^N}\right>
  =
  \left< 0\right|\tilde{X}  \left|0\right>\, ,
\end{equation}
where $\tilde{X}=G^{-1}_{\mathbb{H}_{2^N}} X
G_{\mathbb{H}_{2^N}}$ is the conjugation by the operators implementing 
the conformal mapping. Note that the out-vacuum $\left< 0\right|$ is
invariant under this mapping because $f_N(z)\sim z$ when $|z|\to
\infty$. Inserting the stress tensor in this correlator we have
\begin{equation}
  \begin{split}
  &\left< 0\right|X T(z) \left| \mathbb{H}_{2^N}\right>\\
  &=
  \left( h'_N(w)\right)^{-2}
  \left[
    \left< 0\right|\tilde{X}T(w)  \left|0\right> -
    \frac{c}{12}\{h_N;w\}
    \left< 0\right|X  \left|\mathbb{H}_{2^N}\right>
  \right]\, ,
  \end{split}
\end{equation}
where $\{h_N;w\}$ is the Schwarzian derivative \eqref{eq:schwarz_der}.
We use this starting point to compute
\begin{align}
  \label{eq:TwLn_Hk}
  &\left< 0\right|X \left(L_n-L_{-n} \right) \left| \mathbb{H}_{2^N}\right>
   \\ \label{eq:TwLn_Hk_Tz}
  &=
  \left< 0\right|X 
  \oint_C\frac{{\rm d}z}{2\pi i}z(z^{n}-z^{-n})T(z)\left| 
    \mathbb{H}_{2^N}\right>\\
  &=
  \label{eq:Tw_c}
  \left< 0\right|\tilde{X}
  \oint_{\tilde{C}}\frac{{\rm d}w}{2\pi i}J_{N,n}(w) T(w)\left|0\right>
  \\   \label{eq:Tw_c_2nd} 
  &-
  \frac{c}{12}
  \oint_{\tilde{C}}\frac{{\rm d}w}{2\pi i}J_{N,n}(w) \{g_N;w\}
  \left< 0\right|X  \left|\mathbb{H}_{2^N}\right>\, .
\end{align}
The contour $\tilde{C}$ above encircles the points $w=\pm 2$ 
and lies in the region of validity of the Laurent
expansion \eqref{eq:cmap_uph_Hk}, and we have defined 
\begin{equation}
  \label{eq:JNn}
  J_{N,n}:= \frac{h_N}{h'_N }
  \left(h^n_N-h^{-n}_N\right)\, .
\end{equation}
Further one can check that 
\begin{equation}
  \label{eq:JNn}
  J_{N,n}(w) = (w^2-4) U_{n-1}\left(\smfrac{w}{2} \right)
  + \sum_{p=K-n}^\infty \frac{\alpha'_p}{w^p}\, ,
\end{equation}
where 
\begin{equation}
  \label{eq:Un}
  U_{n-1}\left(\smfrac{w}{2} \right)
  =
  \sum_{k=0}^{ \lfloor \frac{n-1}{2} \rfloor}(-1)^k
  \binom{n-1-k}{k} w^{n-1-2k} 
\end{equation}
is the Chebyshev polynomial of the second kind and $\alpha'_p$ are
opportune coefficients. Note then that as $N\to \infty$ the function
in eq.~\eqref{eq:JNn} reduces to a polynomial of degree $n+1$.  We
focus now on the the integral in equation \eqref{eq:Tw_c}. Computing
the residue at the origin, the polynomial part of $J_{N,n}$ gives
zero, while mapping the contribution of the other non-analytic part back to
the original multi-slit geometry one finds
\begin{equation}
  \label{eq:Tw_int}
  \begin{split}
    &   \left< 0\right|\tilde{X}
  \oint_{\tilde{C}}\frac{{\rm d}w}{2\pi i}J_{N,n}(w) T(w)\left|0\right>\\
  &=\sum_{p=K-n}^\infty \mu_p 
  \left< 0\right|X L_{-p-1}\left|\mathbb{H}_{2^N}\right>\\
  &+ \frac{c}{12}\sum_{p=K-n}^\infty \alpha'_p
  \oint_{\tilde{C}}\frac{{\rm d}w}{2\pi i}w^{-p}\{h_N;w\}
  \left< 0\right|X \left|\mathbb{H}_{2^N}\right>\, ,
  \end{split}
\end{equation}
where $\mu_p$'s are some coefficients whose explicit expression is not needed.
Plugging this back in \eqref{eq:Tw_c} we have
\begin{equation}
  \label{eq:TwLn_Hk_2}
  \begin{split}
  &\left< 0\right|X \left(L_n-L_{-n} \right) \left| \mathbb{H}_{2^N}\right>
  =
  \sum_{p=K-n}^\infty \mu_p 
  \left< 0\right|X L_{-p-1}\left|\mathbb{H}_{2^N}\right>\\
  &-
  \frac{c}{12}
  \oint_{\tilde{C}}\frac{{\rm d}w}{2\pi i} 
  (w^2-4) U_{n-1}\left(\smfrac{w}{2} \right)
  \{h_N;w\}
  \left< 0\right|X  \left|\mathbb{H}_{2^N}\right>\, .
  \end{split}
\end{equation}
We now separate the $N\to\infty$ result from the finite $N$ contributions
in the Schwarzian derivative to get an expression of the following form:
\begin{equation}
  \label{eq:sch_f}
  \{g_N;w\} = \frac{6}{\left(w^2-4\right)^2}+ 
  \sum_{p=K+3}\frac{\alpha''_p}{w^p}\, .
\end{equation}
Both terms in the equation above give a non-vanishing contribution:
\begin{equation}
  \begin{split}
  &-\frac{c}{12}
  \oint_{\tilde{C}}\frac{{\rm d}w}{2\pi i} 
  (w^2-4) U_{n-1}\left(\smfrac{w}{2} \right)
  \{h_N;w\}
  \left< 0\right|X  \left|\mathbb{H}_{2^N}\right>\\
  &=-\frac{nc}{8}(1+(-1)^n)+ A_n  \, ,
  \end{split}
\end{equation}
where we called $A_n$ the number resulting from the the integration of
the second term in \eqref{eq:sch_f}. Clearly one has $A_n=0$ if $n\le K$.
In the end one finds the following gluing condition for the multi-slit geometry:
\begin{equation}
  \label{eq:def_gluing}
  \begin{split}
  &\left(L_n - L_{-n} + \frac{n c}{8} \left[1 + (-1)^n \right] \right)
  \left|\mathbb{H}_{2^N}\right> \\
  &=\left(\sum_{p=K-n}^\infty \mu_p 
   L_{-p-1} + A_n\right) \left|\mathbb{H}_{2^N}\right>  \, .
  \end{split}
\end{equation}
This deformed constraint reduces to the one of
eq.~\eqref{eq:gluing_rect0} for the semi-rectangle when
$N\to \infty$.  This is enough to prove that the boundary state
\eqref{eq:Bstate} obtained as limit of the geometries with $2^N-1$
slits indeed satisfies the gluing condition \eqref{eq:gluing_rect0}.

\subsection{Comparison with partition functions}
\label{sec:comp_part_fun}

A next obvious step is to consider the amplitude associated to the boundary state eq.~(\ref{eq:Bstate})
defined in the previous section and compare it with the known result
for the universal part of the partition function on a rectangle.  We
recall first this result.  Since we consider the case of the same boundary
condition on each side, we note that the partition
function for a rectangle of length $L$ and width $L'$ will have to be
a modular form. If we call the universal part of the partition function
$\Zrec$
\begin{equation}
  Z=e^{f_{\rm b} LL'}e^{f_{\rm s}(L+L')} \Zrec\, ,
\end{equation}
where $f_{\rm b}$ and $f_{\rm s}$ are bulk and surface energies, we have
 \begin{equation}
 Z_{\mathcal{R}}(L,L')=Z_{\mathcal{R}}(L',L)\, .
 \end{equation}

Define as usual $\tau:=i L'/L$, $q:=e^{2\pi i \tau}$ and the Dedekind
eta function as $\eta(\tau) = q^{1/24}\prod_{n=1}^\infty(1-q^n)$.
The partition function $\Zrec$ was computed in \cite{Kleban1991} and up
to possible proportionality coefficients is:
\begin{align}
  \label{eq:rect_Zfree}
  \Zrec &= L^{c/4} \eta(\tau)^{-c/2}\\\nonumber
  &= L^{c/4} 
  q^{-c/48} 
  \left(
    1
    +
    \frac{c}{2}q
    +
    \frac{c (c+6)}{8} q^2 +\dots\right)\, .
\end{align}
As it was observed in \cite{Kleban2003}, this partition function can
be also derived directly from modularity arguments. Indeed, on top of
invariance under modular inversion, as we have discussed the presence
of corners corresponds to an effective weight $c/16$ for each corner,
which fix the non-trivial scaling of the partition function.

Now we can  take the scalar product of the boundary state with itself
to form the amplitude
\begin{align}
  \label{eq:ampli_ish}
  \Arec
  = 
  \begin{tikzpicture}[]
   \useasboundingbox (-0.2,0.2) rectangle (1.1,1);
    \draw (0,0) rectangle (1,0.7); 
    \draw[fill=black] (0,0.7) circle (0.04); 
    \draw[fill=black] (1,0.7) circle (0.04); 
    \draw[fill=black] (0,0) circle (0.04); 
    \draw[fill=black] (1,0) circle (0.04); 
    \draw[->] (0.5,0.1) -- (0.5,0.6); 
  \end{tikzpicture}
  &:= 
  \langle B^o|
  \hat{q}^{L_0-c/24}
  |B^o \rangle\\
  \label{eq:A_cn}
  &=
  \hat{q}^{-c/24}\sum_{n\ge 0}c_n \hat{q}^n
  \, ,
\end{align}
where $\hat{q}:=\sqrt{q}$ is the relevant combination of $L'/L$
appearing in the transfer matrix on a strip, and the arrow is the
direction of imaginary time. The following relation
between this amplitude and the partition function $\Zrec$ should hold
\begin{equation}
  \label{eq:rel_A-Z}
  \Arec=L^{-c/4}\Zrec=\eta(\tau)^{-c/2}\, .
\end{equation}
We have verified this relation by computing the first coefficients
$c_n$ in \eqref{eq:A_cn} using the commutation relations of the
Virasoro algebra and comparing them with the power series of
$\eta^{-c/2}$, confirming the validity of our derivation up to a very
high order ($n=52$) in $\hat{q}$. We note that as expected from the left-right
symmetry of the boundary conditions of the problem, the boundary
states couple only to descendants of the identity of even level
(so that $c_{2n+1}=0$ in eq.~\eqref{eq:A_cn}).

\subsection{Convergence and multi-slit geometry}

In this subsection we discuss further the $N \to \infty$ limit
illustrated in Fig.~\ref{fig:limit}, by computing amplitudes
involving the finitized boundary state
\begin{equation}
 \label{eq:Bnstate}
 \left|\mathbb{H}_{2^N} \right> = e^{- \frac{1}{2^{N-1}} L_{-2^N}}
 \dots e^{-\frac{1}{2} L_{-4}} e^{-L_{-2}} \left| 0\right>
\end{equation}
corresponding to $2^N-1$ slits. We are interested in the generating
functions
\begin{equation}
 P_N(q) = \left( \langle \mathbb{H}_{2^N} 
   | q^{L_0/2} | \mathbb{H}_{2^N} \rangle \right)^{2/c} =
 \sum_{k=0}^\infty p_k^{(N)} q^k \,.
\end{equation}
Since the expected limit is
\begin{equation}
 \lim_{N \to \infty} P_N(q) = q^{1/24} \eta(\tau)^{-1} =
 \prod_{m=1}^\infty \frac{1}{1-q^m} = \sum_{k=0}^\infty p_k q^k \,,
\end{equation}
where $p_k$ is the number of partitions of the integer $k$, we
expect that the $p_k^{(N)}$ will somehow converge to $p_k$. We
shall see below how this occurs.

The case $N=1$ is easily dealt with analytically. Using induction
one can prove that
\begin{equation}
 \langle 0 | L_2^k L_{-2}^k | 0 \rangle =
 \frac{k!}{2^k} \prod_{p=0}^{k-1} (8p+c) \,.
\end{equation}
Developing the exponential $e^{-L_{-2}}$ one then shows that
\begin{equation}
 P_1(q) = (1-4q)^{-1/4} = 1 + q + \frac{5}{2} q^2 + \ldots \,.
\end{equation}
In the next case, $N=2$, a direct computation gives
\begin{equation}
 P_2(q) = 1 + q + 2 q^2 + 3 q^3 + \frac{33}{4} q^4 + \ldots \,.
\end{equation}
By automatizing the computations in the Virasoro algebra using
{\sc Mathematica} we have been able to obtain
the $P_N(q)$ up to order $q^{26}$. In the case of $P_2(q)$
the results are consistent with the conjecture
\begin{equation}
 P_2(q) = \frac{(1+2q)^{1/2} (1+4q^2)^{5/8}}{(1-16q^4)^{3/4}} \,.
\end{equation}

For higher $N$ we have not been able to conjecture---let alone
derive---such exact expressions. We find
\begin{eqnarray}
 P_3(q) &=& 1 + q + 2 q^2 + 3 q^3 + 5 q^4 +
 7 q^5 + 11 q^6 \nonumber \\
 &+& 15 q^7 + \frac{245}{8} q^8 + \ldots +
 \frac{14988511}{512} q^{26} + \ldots
\end{eqnarray}
and
\begin{equation}
 P_4(q) = \sum_{k=0}^{15} p_k q^k + \frac{4005}{16} q^{16} +
 \ldots + \frac{27657}{8} q^{26} + \ldots \,,
\end{equation}
while $P_5(q)$ agrees with $\sum_{k=0}^\infty p_k q^k$ at least up
to order $q^{26}$.

So we observe from these examples that the coefficients $p_k^{(N)}$ are
non-negative rationals that coincide with the integers $p_k$ for $k =
0,1,\ldots,2^N-1$. The first deviating coefficient occurs for $k_0 =
2^N$, and we have then $p_{k_0}^{(N)} > p_{k_0}$.  We conjecture that
these observed properties hold true for any $N \ge 1$.

\section{The case of free theories}
\label{sec:free_theo}

While the expression (\ref{eq:Bstate}) is general, there are more natural ways to think of the boundary state in the case of free theories. There, like when the boundary is a circle instead of a segment, expressions as coherent states over the bosonic/fermionic modes are possible. The comparison with results in the first section is intricate, and leads to remarkable identities. 
We note that  some of the coherent states expressions  presented in this section
 have an important overlap with  \cite{Imamura2006,Imamura2008,Yin2007} \cite{thank_schomerus}.

\subsection{Free boson}
\label{sec:boson}

The simplest case one can deal with is the free boson. We introduce
then harmonic oscillators $a_n, n\in \mathbb{Z}$ satisfying
$[a_m,a_n]=m\delta_{m+n,0}$, $a_{m}|0\rangle=0$ if $m>0$, and
$a_0|0\rangle=0$, with which we represent as usual the Virasoro
generators:
\begin{equation}
  \label{eq:Vir_fb}
  \begin{split}
    L_{n}&=\frac{1}{2}\sum_{m\in\mathbb{Z}} a_{n-m}a_m 
    \, , \quad \mbox{if } n\neq 0\, ,\\
    L_0&=\frac{1}{2}a_0^2+\sum_{m\ge 1} a_{-m}a_m \, .
  \end{split}
\end{equation}

Substituting this expression in \eqref{eq:gluing_rect}, we realize that
the solution should be of the form:
\begin{equation}
  \label{eq:B_fb}
  |B_\phi^o\rangle= \exp\left(-\sum_{n>0}\frac{1}{2n} 
  a_{-n}^2\right)\, .
\end{equation}
To verify this one simply has to use twice the following 
relation
\begin{equation}
  \label{eq:am_a-m_B}
  \left(a_m+a_{-m}\right)|B^o_\phi\rangle=0 \, ,\quad m>0\, .
\end{equation}
The extra term $n/8(1+(-)^n)$ comes in because if $n=2p$ we have
$a_p^2|B^o_\phi\rangle=(a_{-p}^2-p)|B^o_\phi\rangle$.
Another straightforward exercise is computing the amplitude associated  
with this boundary state. We have 
\begin{eqnarray}
  \nonumber \langle B^o_\phi|\hat{q}^{L_0-1/24}|B^o_\phi\rangle
  &=& q^{1/48}\prod_{m>0} \left(\sum_{s\ge 0}\frac{(2s)!}{s!s!}\left(\frac{q^{m}}{4}
    \right)^{s} \right) \\
  &= & \frac{1}{\sqrt{\eta(\tau)}}\, ,
\end{eqnarray}
in agreement with \eqref{eq:rel_A-Z}.
Alternatively one could also express $L_n$
in \eqref{eq:Bstate} using \eqref{eq:Vir_fb} and check that the two
expressions agrees for $c=1$.

Further we note that this result for the partition function of the free
boson can also be obtained by computing the determinant of the
Laplacian with, say, Neumann boundary conditions on all four
sides.  To our knowledge this result appeared first in \cite{DavDup}
(there is factor of $LL'$ there due to the way the zero mode is
subtracted):
\begin{equation}
  \label{eq:laplN}
  \det(-\Delta)
  =
  L^{-1/2} \eta(\tau)\, .
\end{equation}
We note the explicit presence of the anomaly term $L^{c/4}$ in the partition
function given by $\Zrec = (\det(-\Delta))^{-1/2}$.

It is interesting to observe that  eq.~\eqref{eq:laplN} corresponds, using the matrix-tree theorem,  to the partition function
of spanning trees, equivalent in turn to dense polymers or symplectic fermions---all variants of  $c=-2$ CFT. (More generally, partition functions for $c=-2$ with
different boundary conditions on the sides will be given
by taking the power $-2$ of the expressions for the free boson,
since the boundary condition changing operators have dimensions $1/16$
for $c=1$ and $-1/8$ for $c=-2$. This will be exploited in a further paper
\cite{Bondesan2011}.)

\subsection{Majorana fermions}
\label{sec:ising}

We consider now the case of Majorana fermions. We will go 
carefully through the derivation of the gluing condition which
turns out to be more tricky in this case, and then derive the boundary
state by computing the correlator in the rectangular geometry
using a conformal mapping to the upper half plane.

First, we fix notations and conventions.  The fermionic modes are
defined via the expansion of the bulk fermionic fields
\begin{equation}
\psi(z)=\sum_{r\in I} \frac{\psi_r}{ z^{r+1/2}}\, ,
\quad 
\bar{\psi}(\bar{z})=\sum_{r\in I} \frac{\bar{\psi}_r}{\bar{z}^{r+1/2}}
\end{equation}
where $I=\mathbb{Z}$ in the Ramond (R) sector and $I=\mathbb{Z}+1/2$ in
the Neveu-Schwartz (NS) sector.  The OPEs
\begin{equation}
\psi(z)\psi(w)=\frac{1}{z-w}\, ,\quad 
\bar{\psi}(\bar{z})\bar{\psi}(\bar{w})=\frac{1}{\bar{z}-\bar{w}}
\end{equation}
lead to
\begin{equation}
\left\{\psi_r,\psi_s\right\}=
\left\{\bar{\psi}_r,\bar{\psi}_s\right\}=\delta_{r+s,0}\, .
\end{equation}
Let us recall first the condition in the cylinder geometry with
Dirichlet (D) or Neumann (N) boundary conditions on the bottom
\cite{Callan1987}.  For simplicity we restrict to the NS case, where
fermions are periodic in the plane, and thus antiperiodic on the
cylinder.  The cylinder can be obtained from the plane (with complex
coordinate $z$) by the mapping $w=i\ln z$ or $z=e^{-iw}$.  Set
$w=x+iy$. The mode expansion now reads
\begin{align}
\psi(x,y)&=(-i)^{1/2}\sum_{r} \psi_{r}e^{irx-ry}\, ,\\ 
\bar{\psi}(x,y)&=i^{1/2} \sum_r \bar{\psi}_r e^{-irx-ry}\, .
\end{align}
The boundary conditions are then 
\begin{equation}
\label{eq:BC_majo_cyl}
\bar{\psi}(x,y=0)=\epsilon \psi(x,y=0)\, ,
\end{equation}
where $\epsilon=1$ for N and $-1$ for D.
The boundary state associated to these equations is well-known and has
the form of (fermionic) coherent states:
\begin{equation}
|B^p_\psi\rangle
\propto 
\exp\left(\epsilon i
\sum_{p=0}^\infty \psi_{-p-1/2}\bar{\psi}_{-p-1/2}
\right)|O\rangle\, ,
\end{equation}
where 
\begin{equation}
\psi_{p+1/2}|O\rangle=0\, , \quad 
\bar{\psi}_{p+1/2}|O\rangle=0\, , \quad p\in \mathbb{Z}_{\ge 0}\, .
\end{equation}

Now imagine doing the same problem in a geometry rotated by $\pm \pi/2$,
obtained by the mapping $w'=\pm iw$. 
In this rotation, we use the general formulas for transforming Majorana fermions, which are objects of dimension $1/2$, and thus 
$\psi(w')(dw')^{1/2}=\psi(w)(dw)^{1/2}$, and similarly
for  $\bar{\psi}$.
Hence the equations characterizing the boundary state now would  read 
$\bar{\psi}=\pm \epsilon i\psi$ instead of \eqref{eq:BC_majo_cyl}: 
when dealing with fermions, the D or N boundary condition is not represented 
by a unique equation, but depends on the orientation of the boundary.  

Let us go back now to the geometry of interest of
Fig.~\ref{fig:bstate}, and for definiteness choose, say, D boundary
conditions on all sides and $L=\pi$.  The case with N boundary
conditions can be treated in a similar way, and lead in the end to the
same result.  According to our discussion we have
\begin{align}
\bar{\psi}(x=0,y)&=i\psi(x=0,y)\, ,
\quad y\in [-\infty,\infty]\nonumber\\
\bar{\psi}(x=\pi,y)&=-i\psi(x=\pi,y)\, , \quad y\in [-\infty,\infty]
\end{align}

Fully open boundary conditions require identification (up
to monodromy conditions) of left and right modes from both the direct
and crossed channels points of view. It is useful therefore to define
\begin{equation}
  \label{eq:extent}
  \psi(x,y)=-i\bar{\psi}(-x,y)\, ,\quad x\in [-\pi,0]
\end{equation}
so now we can re-express everything in terms of only one type of
fermions---we choose $\psi$. Moreover, by construction, $\psi$ is now
regular at the origin.  By symmetry we can also extend fermions on the
other side of the right boundary by defining
\begin{equation}
\psi(2\pi-x,y)=i\bar{\psi}(x,y)
\end{equation}
so $\psi$ is now regular at $x=\pi$. Using \eqref{eq:extent} gives finally
\begin{equation}
\psi(x+2\pi)=-\psi(x)\, , \quad x\in [-\pi,\pi]
\end{equation}
so that we now have a single species of fermions defined on the circle
$[-\pi,\pi]$, with NS (antiperiodic) boundary conditions.

All the manipulations so far are there to handle the boundary conditions at
 $x=0$ and $x=\pi$. 
We have now, on top of this, to handle the boundary condition at the bottom 
of the system, that is for $y=0$. Taking $\epsilon=-1$ 
in \eqref{eq:BC_majo_cyl}, from \eqref{eq:extent} we have
\begin{equation}
\left[\psi(x,y=0)+i\psi(-x,y=0)\right]
|B^o_{\psi}\rangle=0\, , \quad x\in [0,\pi]
\end{equation}
Going through the definitions carefully shows that the relative sign of 
the two fermion terms in this equation switches in the interval 
$x\in [-\pi,0]$. This is in line with the fact that fermion correlators 
change sign under reflection. So the final equation has to be for 
$x\in [-\pi,\pi]$
\begin{equation}
\left[\psi(x,y=0)+\hbox{sign}(x)i\psi(-x,y=0)\right]
|B^o_{\psi}\rangle=0\, .
\end{equation}
The sign function introduces serious complications with respect to the
case of ordinary boundary states. Notice that it disappears when
considering conditions for the Virasoro generators, since the stress-energy
tensor is quadratic in the fermions.
We represent the sign function through  the Fourier series
\begin{equation}
\hbox{sign}(x)=\sum_{m=-\infty}^\infty a_m e^{imx}
\end{equation}
with 
\begin{align}
a_{2m}&=0\nonumber\\
a_{2m+1}&=-\frac{2i}{(2m+1)\pi}\, .
\end{align}
We can introduce the matrix $A$ defined through
\begin{align}
A_{m+1/2,n+1/2}&=\frac{1-(-1)^{m+n+1}}{\pi(m+n+1)},~~~m+n+1\neq 0\nonumber\\
A_{m+1/2,-m-1/2}&=0
\end{align}
so the equation satisfied by the boundary state reads, in terms of the
fermionic modes
\begin{equation}
\label{eq:def}
\left(\psi_{m+1/2}+\sum_n A_{m+1/2,n+1/2}\psi_{n+1/2}\right)|B^o_{\psi}\rangle=0\, .
\end{equation}
Now decompose the $A$ matrix (whose labels are half odd integers) into
four blocks corresponding to the label signs:
\begin{equation}
A=\left(\begin{array}{cccc}
a_{m,n}& b_{m,n}\\
-b_{m,n}&-a_{m,n}\end{array}\right)\, ,
\end{equation}
where we have defined blocks $a_{m,n}\equiv
A_{m+1/2,n+1/2}$, and $b_{m,n}\equiv A_{m+1/2,-n-1/2}$, with $m,n\geq 0$.
Take now $m\geq 0$ (for $m<0$ one gets an equivalent condition) 
and consider eq.~\eqref{eq:def}, which we can rewrite as
$$
\left(\psi_{m+1/2}+\sum_{n\geq 0}  a_{mn}\psi_{n+1/2}+b_{mn}\psi_{-n-1/2}\right)
|B^o_{\psi}\rangle=0\, .
$$
Collecting the positive and negative fermion modes and inverting the
system gives the equation
$$
\left(\psi_{m+1/2}+\sum_{n\geq 0}\left[(1+a)^{-1}b \right]_{mn}
\psi_{-n-1/2}\right)|B^o_{\psi}\rangle=0\, .
$$
Now set $G:=-(1+a)^{-1}b=-G^{T}$. 
The solution of this system is (up to proportionality coefficients)
 the coherent state
\begin{equation}
\label{eq:majo_coherent}
|B_{\psi}^o\rangle
= 
\exp\left(
\sum_{0\leq m< n}^\infty G_{m,n}\psi_{-m-1/2}\psi_{-n-1/2}
\right)|O\rangle \, .
\end{equation}

With this approach the matrix $G_{m,n}$ is somewhat undetermined.
We present then a different strategy to compute it using 
correlation functions. As in section \ref{sec:BstateSLE} we define
the boundary state in the geometry denoted by ${\cal D}$ 
(fig.~\ref{fig:map_D_UHP} (left)),
and consider the mapping to the upper half plane 
$\mathbb{H}$ (fig.~\ref{fig:map_D_UHP} (right)),
viz.~$w=z+z^{-1}$.
We thus consider D boundary conditions on all sides of the boundary of
${\cal D}$, so that $|B^o_{\psi}\rangle$ should be the state
representing this boundary.  Calling $|O\rangle$ the ground state of
the model at infinity in $\mathbb{H}$ we have
\begin{align}
\label{eq:corr_D_H}
\langle \psi(z_1)\psi(z_2)\rangle_{{\cal D}}
&= \langle O|\psi(z_1)\psi(z_2)|B^o_{\psi}\rangle\\
&=\left(\frac{\partial w_1}{\partial z_1}\frac{\partial w_2}{\partial z_2}
\right)^{1/2}\langle\psi(w_1)\psi(w_2)\rangle_{\mathbb{H}}\, ,
\end{align}
where on the rhs the correlator is evaluated in $\mathbb{H}$ with the usual 
D boundary conditions. Using that this correlator must simply be 
$1/(w_1-w_2)$ (since it involves only right movers, which are not affected 
by the boundary), and evaluating derivatives gives straightforwardly that 
$$
\langle O|\psi(z_1)\psi(z_2)|B_{\psi}^o\rangle
=\frac{1}{z_1-z_2}
\frac{\sqrt{1-\frac{1}{z_1^2}}\sqrt{1-\frac{1}{z_2^2}}}{1-\frac{1}{z_1z_2}}\, .
$$
From the result (\ref{eq:majo_coherent}) we can write the boundary state as
\begin{equation}
|B_{\psi}^o\rangle
=\;
:\exp\left(\oint \frac{dz}{2i\pi}\oint \frac{dz'}{2i\pi}
\psi(z)G(z,z')\psi(z')\right):|O\rangle\, ,
\end{equation}
where $G(z,z')$ corresponds to the generating function of the numbers
$G_{m,n}$ introduced above. The left hand side of
eq.~\eqref{eq:corr_D_H} can now be evaluated using Wick's theorem. We find
in the end that
\begin{equation}
  G(z_1,z_2)=
  \frac{1}{2(z_2-z_1)}
  \left(\frac{\sqrt{1-\frac{1}{z_1^2}}\sqrt{1-\frac{1}{z_2^2}}}
    {1-\frac{1}{z_1z_2}}-1\right)\, .
\end{equation}
We need to evaluate this expression in the domain $|z_1|,|z_2|>1$
(because of the geometry of our problem), and expand it as
\begin{equation}
G(z_1,z_2)= \frac{1}{2z_1z_2}\sum_{m,n=0}^\infty \frac{G_{mn}}{z_1^{m}z_2^{n}}\, .
\end{equation}
(Observe that on top of $G_{mn}=-G_{nm}$, we have $G_{mn}=0$ if $m+n$ is even.)
As a result,  we now  have the generating function for the the quadratic form appearing
in \eqref{eq:majo_coherent}. The first few values of $G_{m,n}$ read:
\begin{eqnarray*}
G_{01}=\frac{1}{2}\\
G_{03}=\frac{1}{8},G_{12}=\frac{5}{8}\\
G_{05}=\frac{1}{16},G_{14}=\frac{3}{16},G_{23}=\frac{5}{8}\\
G_{07}=\frac{5}{128},G_{16}=\frac{13}{128},G_{25}=\frac{25}{128},
G_{34}=\frac{81}{128}
\end{eqnarray*}

\vspace{0.25cm}

We can then compare the result for the boundary state found in this
section with the specialization at $c=1/2$ of formula \eqref{eq:Bstate}
by expressing the Virasoro modes in a standard way in terms of
fermionic ones:
\begin{equation}
\label{eq:Lnfermions}
  L_n 
  = \frac{1}{2} \sum_{k\;\in\;\mathbb{Z}+\frac{1}{2}}k :\psi_{-k+n}\psi_{k}: \, ,
  \quad n\in \mathbb{Z}\, .
\end{equation}
We have verified the agreement for the first few descendants up to level
$8$ in the Virasoro modes of the two derivations for the Ising CFT.

Further by comparing the amplitudes of boundary states we find:
\begin{equation}
  \label{exp1}
  \begin{split}
    &\langle B^o_{\psi}|\hat{q}^{L_0-1/48}|B^o_{\psi}\rangle =\\
    &=
    q^{-1/24}\left(
      1+{q^2\over 4}+{13\over 32}q^4+{55\over 128}q^6+{1235\over 2048}q^8+\dots
    \right)
    \, ,
  \end{split}
\end{equation}
where $|B^o_{\psi}\rangle$ as in eq.~\eqref{eq:majo_coherent} and $L_0$ is
expressed in terms of fermions using eq.~\eqref{eq:Lnfermions}. Of course, we know on the other hand from the general case \eqref{eq:rel_A-Z} that 
\begin{equation}
\label{exp2}
\langle B^o_{\psi}|\hat{q}^{L_0-1/48}|B^o_{\psi}\rangle
=[\eta(\tau)]^{-1/4}\, .
\end{equation}
Matching (\ref{exp1}) and (\ref{exp2}) gives rise to intriguing 
combinatorial identities, which we have of course checked to high order, but are not able to prove.

\section{Lattice models}

\subsection{Lattice discretization and numerics}
\label{sec:numerics_loops}

We propose now a lattice discretization of the boundary state and
verify numerically the contribution of the first levels
descendants of the identity in \eqref{eq:Bstate}
by computing the scaling limit of scalar products in the
lattice model.  

The lattice model we consider is the dense loop model based on the
adjoint representation of the Temperley-Lieb algebra acting on $N$ (even)
strands TL${}_{N}(\beta)$ ($\beta$ is the so-called weight of loops),
and imposing free boundary conditions at both boundaries. At the
critical point, the anisotropic version of the model is defined by the
Hamiltonian:
\begin{equation}
  \label{eq:H_TL}
  H = -\sum_{i=0}^{N-2}e_i\, ,
\end{equation}
where $e_i$ are the TL generators. This Hamiltonian acts on link
states $|s\rangle_N$, reduced TL diagrams keeping track of connectivities
of sites.  The Hamiltonian can be put in a
triangular block form, with $j=0,1,\dots,N$, the number of through lines,
 indexing each block \cite{Martin1991}. On the lattice we
define the transpose ${}_N\langle s|$ of a state by turning it upside
down and the bilinear form which we call loop scalar product
${}_N\langle s_1| s_2\rangle_N$, associating to the link states the
diagram obtained by gluing $|s_2\rangle_N$ with ${}_N\langle
s_1|$. The result is given by $\beta^n$, with $n$ the number of loops 
thus formed, and we get zero if we contract two strings.  For instance we
have
\begin{equation}
\label{loopscalar}
\begin{tikzpicture}
\begin{scope}[xshift=0cm]
        \draw (0,0) node{$\left< {\; \begin{tikzpicture}
        \draw[thick] (0,0) arc (-180:0:0.1 and 0.25);
        \draw[thick] (0.4,0) arc (-180:0:0.3 and 0.25);
        \draw[thick] (0.6,0) arc (-180:0:0.1 and 0.15);
        \end{tikzpicture}}
        \; | \; {\begin{tikzpicture}
        \draw[thick] (0,0) arc (-180:0:0.5 and 0.25);
        \draw[thick] (0.2,0) arc (-180:0:0.1 and 0.15);
        \draw[thick] (0.6,0) arc (-180:0:0.1 and 0.15);
        \end{tikzpicture}} \; \right>$};
\end{scope}
\draw (1.75,-0.03) node{$=$};
\begin{scope}[xshift=2.15cm]
        \draw[thick] (0,0) arc (180:0:0.1 and 0.25);
        \draw[thick] (0.4,0) arc (180:0:0.3 and 0.25);
        \draw[thick] (0.6,0) arc (180:0:0.1 and 0.15);
        \draw[thick] (0,0) arc (-180:0:0.5 and 0.25);
        \draw[thick] (0.2,0) arc (-180:0:0.1 and 0.15);
        \draw[thick] (0.6,0) arc (-180:0:0.1 and 0.15);
\end{scope}
\draw (3.75,0.05) node {$= \; \beta^2\, .$};
\end{tikzpicture}
\end{equation}
This is the usual bilinear form used for the TL algebra \cite{Jones2006}.
For generic $\beta$ the loop scalar product is not positive definite.
There is no reason to expect otherwise, as the theories we are dealing
with are not unitary.  The continuum limit of this loop
model, when we parametrize the loop weight as $\beta=2\cos(\pi/(p+1))$ ($p \geq 1$ is a real parameter), 
is a CFT with central charge
$c=1-6/(p(p+1))$. 
Further the bilinear form introduced above
flows in the continuum limit  to the Virasoro bilinear form, as already
observed in \cite{Dubail2010}. This will allow us to measure the
boundary state on the lattice.
We claim that the lattice state renormalizing in the scaling
limit to our boundary state introduced above is the following link
pattern
\begin{equation}
  \label{eq:scaling_state_lattice}
  \begin{tikzpicture}[scale=0.5]
    \node at (-3,-0.15) {$|B^o\rangle_N:=\frac{1}{\sqrt{\beta^N}}$};
    \begin{scope}[xshift=0cm]
      \draw (1,0) node{$\left| \; {\begin{tikzpicture}
              \draw[thick] (0,0) arc (-180:0:0.1 and 0.15);
              \node at (0.65,0) {$\dots$};
              \draw[thick] (1,0) arc (-180:0:0.1 and 0.15);
            \end{tikzpicture}} \; \right>$};
    \end{scope}
    \draw[->] (4,0)--(6,0);
    \node at (5,-0.5) {$N\to \infty$};
    \node at (8,0) {$\alpha |B^o\rangle\, ,$};
  \end{tikzpicture}
{}  
\end{equation}
where in the continuum limit procedure we discard non-universal
contributions, and introduce the proportionality constant $\alpha$.
Note that the lattice state is normalized according
to the form ${}_N\langle \cdot|\cdot\rangle_N$ and that
only the sector $j=0$ without through lines couples to this state.

To check this relation we have computed numerically the scaling limits
of lattice scalar products ${}_N\langle B^o | k\rangle_N$ of the
boundary state with the $k$-th eigenvector in the sector $j=0$,
ordered according to its energy ($|0\rangle_N$ is the ground state).  
The vectors $|k\rangle_N$ are normalized to $1$ 
using the loop scalar product. In
the $j=0$ sector, the continuum theory has field content given
generically by the quotient of Virasoro Verma modules
$V_{1,1}/V_{1,-1}$ (with the usual Kac table notation $V_{r,s}$ for Verma modules) \cite{Read2007}, and 
the scaling limit $|k\rangle$ of $|k\rangle_N$ will be given 
by a combination of Virasoro descendants at a level 
determined by the energy of $|k\rangle$. We identify then obviously
$|k=0\rangle$ as the vacuum $|0\rangle$ and (due to 
normalization) $|1\rangle=\sqrt{2/c}L_{-2}|0\rangle$,
$|2\rangle = \sqrt{1/2c} L_{-3}|0\rangle$. Unfortunately
we do not know a priori 
which combination of Virasoro descendants contributes to 
$|k\rangle$, for $k>2$.
We however do know the number of descendants at each level, since the
above quotient amounts to setting $L_{-1}|0\rangle = 0$ in the generic
Verma module. 

Due to the anomaly at the corners, we expect the following scaling form
of the lattice scalar products:
\begin{equation}
  \label{eq:mlogBk}
  -\log( {}_N\langle B^o | k \rangle_N )
  = 
  a_0 N
  +
  a_1 \log N
  +
  a_2
  +
  \frac{a_3}{N}
  +
  \frac{a_4}{N^2}
  +
  O\left(\frac{1}{N^3}\right) \, ,
\end{equation}
with
\begin{equation}
  \label{eq:a1}
  a_1 = -\frac{c}{8}\, .
\end{equation}
Note the difference between formula \eqref{eq:mlogBk} and the one for
the scaling in the case of periodic boundary states $|B^p\rangle$, for
which there is no logarithmic term and excited states corresponding to
descendant fields decouple.

In the following we take anyway $a_1$ as a free parameter and fit our
numerical results with \eqref{eq:mlogBk} where we drop the term
$O(N^{-3})$.  $\langle B^o | k \rangle$ can then be extracted from $a_2 = -
\log( \langle B^o | k \rangle/\alpha )$.  
As usual we cannot determine the sign of the scalar product,
which however depends on an arbitrary phase of $\ket{k}_N$. 
The numerical results are presented in table \ref{tab:num_loops}.
\begin{table}[t]
\begin{tabular}{c|c|c|c|}
\cline{2-4}
& $p$ &
numerics &
CFT \\
\hline
\multicolumn{1}{|c}{\multirow{4}{*}{$a_1$}} &
\multicolumn{1}{|c|}{$3$} & $-0.06238\pm 0.00137$ & $-0.0625$ \\
\multicolumn{1}{|c}{}                        &
\multicolumn{1}{|c|}{$4$} & $-0.08659\pm 0.00214$ & $-0.0875$ \\
\multicolumn{1}{|c}{}                        &
\multicolumn{1}{|c|}{$5$} & $-0.09515\pm 0.00033$ & $-0.1$     \\
\multicolumn{1}{|c}{}                        &
\multicolumn{1}{|c|}{$\infty$} & $-0.11706\pm 0.00762$ & $-0.125$     \\
\hline
\multicolumn{1}{|c}{\multirow{4}{*}{$\alpha$}} &
\multicolumn{1}{|c|}{$3$} & $0.97642\pm 0.00447$ & \\
\multicolumn{1}{|c}{}                        &
\multicolumn{1}{|c|}{$4$} & $0.94890\pm 0.00679$ & \\
\multicolumn{1}{|c}{}                        &
\multicolumn{1}{|c|}{$5$} & $0.94000\pm 0.00010$ & \\
\multicolumn{1}{|c}{}                        &
\multicolumn{1}{|c|}{$\infty$} & $0.88731\pm 0.00327$ & \\
\hline
\hline
\multicolumn{1}{|c}{\multirow{4}{*}{$\langle B^o|1\rangle$}} &
\multicolumn{1}{|c|}{$3$} & $0.49994\pm 0.00229$ & $0.5$    \\
\multicolumn{1}{|c}{}                        &
\multicolumn{1}{|c|}{$4$} & $0.58652\pm 0.00422$ &  $\approx 0.591608$    \\
\multicolumn{1}{|c}{}                        &
\multicolumn{1}{|c|}{$5$} & $0.61994\pm 0.00067$ &  $\approx 0.632456$  \\
\multicolumn{1}{|c}{}                        &
\multicolumn{1}{|c|}{$\infty$} & $0.65330\pm 0.01789$ & $\approx 0.707107$  \\
\hline
\multicolumn{1}{|c}{\multirow{4}{*}{$\langle B^o|2\rangle$}} &
\multicolumn{1}{|c|}{$3$} & $0$ &   \multirow{4}{*}{$0$} \\
\multicolumn{1}{|c}{}                        &
\multicolumn{1}{|c|}{$4$} & $0$ &     \\
\multicolumn{1}{|c}{}                        &
\multicolumn{1}{|c|}{$5$} & $0$ &   \\
\multicolumn{1}{|c}{}                        &
\multicolumn{1}{|c|}{$\infty$} & $0$ & \\
\hline
\end{tabular}
\caption{Numerical results for the loop model obtained by 
  fitting the lattice scalar products with \eqref{eq:mlogBk}
  for system sizes $N=8\to 24$.}
\label{tab:num_loops}
\end{table}

The value of $a_1$ and $\alpha$ are obtained from the scaling of
$\langle B^o|0\rangle$. We get good agreement of the numerical result for $a_1$
with the CFT prediction when $p=3$, and the discrepancy increases with $p$.
 The constant $\alpha$ clearly depends on $p$, but we have
not been able to determine if it is a universal quantity or
not. Simulations for the critical $XY$ spin chain in a transverse
field show that it is independent of the parameter $r$ tuning from
$c=1/2$ to $c=1$, suggesting that it could be universal. Further it is
shown in \cite{Bondesan2011} that the ratio of two proportionality
constants $\alpha$ for lattice boundary states corresponding to
different boundary conditions is universal, and so we believe that $\alpha$
should contain a universal part.  
For computing $\langle B^o|k\rangle$, we actually fit 
\begin{equation}
  \label{eq:eq:mlogBk_div}
  -\log\left( \frac{{}_N\langle B^o | k \rangle_N}{{}_N\langle B^o|0\rangle_N}
    \right)
  =
  a_2 +
  \frac{a_3}{N}
  +
  \frac{a_4}{N^2}
  +
  \frac{a_5}{N^3}\, ,
\end{equation}
and drop first points.  According to the identification
$|1\rangle=\sqrt{2/c}L_{-2}|0\rangle$, the numbers $\langle
B^o|1\rangle$ in the column CFT are then $\sqrt{c/2}$.  As for 
the comparison of $a_1$, when
 $p=3$ the agreement is very good,
 while it gets worse for greater values of $p$.
$|2\rangle$ is a state with
conformal dimension $3$, so that our formula of the boundary state
predicts its decoupling.  Remarkably, $\langle B^o|2\rangle$ is exactly
zero within numerical precision for sizes $N\ge 10$.  For higher
excited states numerical simulations do not give results accurate enough
to allow to clearly identify the combinations of Virasoro 
descendants to which they correspond, so that we cannot
compare further the prediction of our formula for the boundary 
state \eqref{eq:Bstate}.
Finally, we note that for $p=1,2$ corresponding respectively to dense
polymers ($c=-2$) and percolation ($c=0$), the presence of null
vectors in the module of the identity require smart tricks for
computing scalar products, such as those employed in
\cite{Dubail2010}, and we have not studied numerically these
cases.

\subsection{Ising chain}
\label{sec:ising_chain}

In this section we give evidence that the fermionic coherent
state of eq.~\eqref{eq:majo_coherent} is the universal part of
the continuum limit of a lattice state expressed in terms of Ising spins.
Similar computations appear in \cite{Stephan2010}.
D boundary conditions considered in the previous derivation correspond to 
fixed boundary conditions for the spins. As we have 
already observed, result \eqref{eq:majo_coherent} holds more 
generally for homogeneous boundary conditions, and here we will consider
instead free boundary conditions (as done for the loop model in section
\ref{sec:numerics_loops}).

We consider the Hamiltonian limit of the $2$D critical
Ising model on a strip with free/free boundary conditions:
\begin{equation}
  \label{eq:H_ising}
  H 
  = 
  -\frac{1}{2} 
  \left( 
  \sum_{i=1}^N \sigma_i^z
  +
  \sum_{i=1}^{N-1} \sigma_i^x \sigma_{i+1}^x
  \right) \, ,
\end{equation}
acting on an Hilbert space made of $N$ (even) copies of the
fundamental representation of su$(2)$.  This model has been
extensively studied in the past (see for example \cite{Lieb1961}).
After briefly reviewing the solution, we will compute the limit of
lattice scalar products which we will compare with the CFT predictions.

We introduce the fermions $c_i,c_i^\dagger$ and perform the
Jordan-Wigner transformation (redefining $\sigma_i^z$ as $-\sigma_i^z$)
\begin{align}
  \label{eq:jw}
  \sigma_i^z = 1 - 2 c_i^\dagger c_i \, ,\quad
  \sigma_i^x = \prod_{j<i}(1 - 2 c_j^\dagger c_j)(c_i + c_i^\dagger)\, ,
\end{align}
to obtain an expression of $H$ quadratic in $c_i^\dagger$:
\begin{equation}
  \label{eq:H_ising_ff}
  H 
  = 
  \sum_{i=1}^N 2 c_i^\dagger c_i
  -
  \frac{1}{2}
  \sum_{i=1}^{N-1} c_i^\dagger c_{i+1} + c_{i+1}^\dagger c_{i}
  +
  c_i^\dagger c^\dagger_{i+1} 
  +
  c_{i+1}c_{i} \, ,
\end{equation}
which can readily be diagonalized through a canonical transformation
\begin{align}
  \psi_{k} \pm \psi^\dagger_{k} 
  = 
  \sum_{i=1}^N 
  \phi^\pm_{ki} 
  \left( 
  c_i  \pm c^\dagger_{i} 
  \right) \, .
\end{align}
We have
\begin{equation}
  \label{eq:H_ising_ff_eta}
  H 
  = 
  \sum_{k=1}^N
  \Lambda_{k}
  \left(
  \psi_k^\dagger \psi_{k} 
  -
  \frac{1}{2}
  \right)
\end{equation}
with 
\begin{align}
  \label{eq:eigen}
  \Lambda_{k} &= 2 \sin\left( \frac{2k-1}{2(2N+1)} \pi \right)\\
  \label{eq:phi}
  \phi^+_{ki} &= (-)^i \frac{2}{\sqrt{2N+1}} 
  \cos\left(\frac{2k-1}{2N+1} \pi \left(i - \frac{1}{2} \right) \right)\\
  \label{eq:psi}
  \phi^-_{ki} &= (-)^{i+1} \frac{2}{\sqrt{2N+1}} 
  \sin\left(\frac{2k-1}{2N+1} \pi i\right) \, .
\end{align}
Note that $\phi^\pm$ are orthogonal matrices.
The ground state of the Hamiltonian is $\gs$, with $\psi_k \gs = 0$ for
every $k$. When $N$ is large, we have
\begin{equation}
  \label{eq:H_N_infty}
  \sum_{k=1}^N\Lambda_k \psi^\dagger_{k}\psi_{k} 
  \simeq 
  \frac{\pi}{N}\sum_{k=1}^N \left(k-\frac{1}{2}\right) \psi^\dagger_{k}\psi_{k} 
  \, .
\end{equation}
If now we introduce the notation $\psi_{-k} \equiv \psi_k^\dagger$ for
$k > 0$, and redefine $\psi_k \to \psi_{k - 1/2}$, we
identify $H$ with $L_0$, where $L_0$ is defined in terms of rescaled
fermions as in \eqref{eq:Lnfermions}.  We recover then that the system in
the continuum limit is described by Ising CFT. For the free
boundary conditions chosen the states propagating in the strip are the
descendants of the identity (sector with an even number of fermions)
and the energy $\psi$ (sector with an odd number of fermions).

Now recall that the Hamiltonian \eqref{eq:H_ising} is related to the
$2$D Ising model in the $\sigma^x$ basis.  Then to make the contact
with the $2$D model, we should rotate $90$ degrees clockwise the spins
of the chain in the $x-z$ plane, that is, we define the free boundary state
$|B_{\psi}^o\rangle_{N}$ as
\begin{align}
  \label{eq:freeN}
  |B_{\psi}^o\rangle_{N} 
  &= 
  \frac{1}{\sqrt{2^{N}}}
  \sum_{\{\mu_i^x = \rightarrow,\leftarrow \}} \ket{\mu_1^x \dots \mu_N^x} 
  = \ket{\uparrow \cdots \uparrow} \, .
\end{align}

We want to compute ${}_N\langle B^o_{\psi}|k\rangle_N$, where $|k\rangle_N$ is the
$k$-th excited state.  One way to deal with that is writing the
projector onto $|B^o_{\psi}\rangle_{N}$ in terms of $c_i$'s (the same trick was used for
example in \cite{Stephan2009})
\begin{equation}
  |B_{\psi}^o\rangle_{N} {}_N\langle B^o_{\psi}|  
  = \prod_{i=1}^N \left( \frac{1 - \sigma_i^z}{2} \right)
  = \prod_{i=1}^N c_i^\dagger c_i \, ,
\end{equation}
so that $|{}_N\langle B_{\psi}^o|k\rangle_N|^2$ 
can be evaluated efficiently using
Wick's theorem.  
Let us start dealing with the ground state. Rearranging the factors of
$c_i^\dagger$ and $c_i$ in a convenient way, we have to compute
\begin{align}
  \gsb \prod_{i=1}^N c_i^\dagger \prod_{i=1}^N c_i \gs 
  = \Pf 
  \left( \begin{array}{cc}
      C & D \\
      -D & -C \end{array} \right)\, ,
\end{align}
where (see also \cite{AbanovToeplitz})
\begin{align}
 C_{ij} &= - C_{ji} = \langle c_i^\dagger c_j^\dagger \rangle = 
\frac{G_{ij} - G_{ji}}{4}\\
D_{ij} &= D_{ji} = \langle c_i^\dagger c_j \rangle = 
\frac{2\delta_{ij} - G_{ij} - G_{ji}}{4}
\end{align}
and we have introduced the function
\begin{equation}
  \label{eq:Gij}
  G_{ij} 
  =
  - \sum_{k=1}^N \phi^-_{ki}\phi^+_{kj} 
   \, .
\end{equation}
Computing the Pfaffian from the square of the determinant and using
properties of block matrices, we have
\begin{equation}
  \label{eq:free_gs_N}
  |{}_N\langle B^o_{\psi}|k\rangle_N|^2
  =
  \det_{1\le i,j \le N}\left(\frac{\delta_{ij}+G_{ij}}{2}\right)\, .
\end{equation}
The scalar product with the excited state $\psi_{k_1}\cdots \psi_{k_s}
\gs$ can be computed by regarding this state as the vacuum state of a
new set of $\psi$'s, where $\psi_k$ and $\psi_k^\dagger$ are
interchanged for $k_1, \dots , k_s$ \cite{Lieb1961}. For these $k$'s
we have $\Lambda_k \to -\Lambda_k$ and $\phi^-_{ki} \to -\phi^-_{ki}$, so
that \eqref{eq:free_gs_N} holds also for $\psi_{k_1}\cdots \psi_{k_s}
\gs$ if we replace $G_{ij}$ with
\begin{equation}
  \label{eq:Gij_exc}
  G_{ij}(k_1,\dots,k_s) = -\sum_{k \, \mbox{\scriptsize{unexc.}}}
  \phi^-_{ki}\phi^+_{kj} + \sum_{k \, \mbox{\scriptsize{exc.}}}
  \phi^-_{ki}\phi^+_{kj}\, .
\end{equation}

We have not been able to find a closed formula for the scalar
products, but eq.~\eqref{eq:free_gs_N} allows for very efficient
numerical computations of the scalar products.  We have fitted results
for $N = 2 \to 500$ (dropping the first points for higher excited
states), using the fit function \eqref{eq:mlogBk}.  The value
$a_1=-1/16$ is always found with a precision of $10^{-6}$ and for the
proportionality constant $\alpha$ we find with a high precision
$\alpha = 1.01937$.  For ${}_N\langle B^o_{\psi}|k\rangle_N$ we have the
results in table \ref{tab:scal_prod_free_scal}.
\begin{table}[h!c]
  \centering
  \begin{tabular}{|c|c|c|c|}
    \cline{2-4}
    \multicolumn{1}{r|}{} 
    & $h_k$ & numerics & CFT \\
    \hline
    $\langle B^o_{\psi}|1\rangle$ & $1/2$ & $0$ & $0$ \\
    \hline
    $\langle B^o_{\psi}|2\rangle$ & $3/2$ & $0$ & $0$ \\
    \hline
    $\langle B^o_{\psi}|3\rangle$ & $2$ & $0.499994 \pm 0.000003$ & $1/2$ \\
    \hline
    $\langle B^o_{\psi}|4\rangle$ & $5/2$ & $0$ & $0$ \\
    \hline
    $\langle B^o_{\psi}|5\rangle$ & $3$ & $0$ & $0$ \\
    \hline
    $\langle B^o_{\psi}|6\rangle$ & $7/2$ & $0$ & $0$ \\
    \hline
    $\langle B^o_{\psi}|7\rangle$ & $4$ & $0.124995 \pm 0.000003$ & $1/8=0.125$ \\
    \hline
    $\langle B^o_{\psi}|8\rangle$ & $4$ & $0.624989 \pm 0.000003$ & $5/8=0.625$ \\
    \hline
    $\langle B^o_{\psi}|9\rangle$ & $9/2$ & $0$ & $0$ \\
    \hline
    $\langle B^o_{\psi}|10\rangle$ & $9/2$ & $0$ & $0$ \\
    \hline
  \end{tabular}
  \caption{Scalar product of $|B^o_{\psi}\rangle$ and $|k\rangle$ from finite size
    scaling (numerics) and comparison with CFT prediction. $h_k$
    is the conformal dimension of the field $|k\rangle$.}
 \label{tab:scal_prod_free_scal}
\end{table}
We note that obviously the sector of the energy $\psi$ ($h_k$
half integer) decouples.
The agreement is very good. It remains intriguing of course, that
\eqref{exp2} does not seem to be easily obtainable from the explicit solution of the lattice model. This is in contrast with similar expressions for other geometries, such as the torus. 

\section{Conclusion}

This general exploration of the boundary states for theories defined on a segment will be used in our next paper to discuss in particular properties of geometrical problems on a rectangle. Of course, the set-up can find many other applications as well, in condensed matter or string theory \cite{Imamura2006,Imamura2008,Yin2007,Cantcheff2008}. Another
field where our boundary state (\ref{eq:Bstate}) might be relevant is
the calculation of quantum information quantities such as entanglement entropies
or overlaps in 1d or 2d systems. For example, in some 2d conformally invariant
wave functions \cite{FradkinMoore,Stephan2009}, the calculation of the entanglement
entropy boils down to the study of expansions like (\ref{eq:mlogBk}). Some references
\cite{Zaletel2011,Stephan2011} have focused on the universal logarithmic term $a_1 \log L$ in (\ref{eq:mlogBk}), but we have shown that the next term $a_2$ contains also a universal piece, and it would be interesting to extend their analysis using our present results.

\subsection{Note added}
After the completion of this work, we became aware of a similar result
published in \cite{Rastelli2001,Ellwood2001,Schnabl2003}, which is
consistent with our main formula eq.~\eqref{eq:Bstate}.

\subsection{Acknowledgements}
R.~B.~and J.~D.~thank J.-M.~St\'ephan for extremely useful discussions and comments, and for carefully reading the manuscript. 
The work of J.~L.~J.  and H.~S.  was supported by a grant from the ANR Projet 2010 Blanc SIMI 4: DIME. JD is supported by a Yale Postdoctoral Prize Fellowship. We thank the IHP, Paris, for its hospitality.

\end{document}